\documentclass[aps,preprint,fleqn,prb,byrevtex]{revtex4}

\usepackage{psfrag,graphicx,here}

\begin{document}  

\title{General calculation of $4f-5d$ transition rates for rare-earth
  ions using many-body perturbation theory}


\newcommand{\Bra}[1]{\left \langle #1 \right |} 
\newcommand{\Ket}[1]{\left | #1 \right \rangle}  
\newcommand{\BraKet}[2]{ \langle #1 | #2  \rangle} 
\newcommand{\ME}[3]{\Bra{#1} #2 \Ket{#3}} 
\newcommand{\RME}[3] {\langle #1 || #2 || #3  \rangle}  
\newcommand{\VectNorm}[1]{ \langle #1 | #1  \rangle>}  
\newcommand{\Project}[1]{\Ket{#1}\Bra{#1}}
\newcommand{\be}{\begin{equation}}
\newcommand{\ee}{\end{equation}}
\newcommand{\bra}[1]{\langle #1 |}
\newcommand{\ket}[1]{| #1 \rangle}
\newcommand{\bk}[2]{\bra{#1} #2 \rangle}
\newcommand{\bok}[3]{\bra{#1} #2 \ket{#3}}
\newcommand{\rme}[3]{\langle #1 || #2 || #3 \rangle}
\newcommand{\rmet}[3]{\langle #3 || #2 || #1 \rangle}
\newcommand{\bnk}[1]{\langle #1 \rangle}
\newcommand{\kb}[2]{\ket{#1} \bra{#2}}
\newcommand{\kk}[1]{\kb{#1}{#1}}
\newcommand{\Sum}[1]{\sum\limits_{#1}}
\newcommand{\Sumb}[1]{\sum'\limits_{#1}}
\newcommand{\ep}{\epsilon}
\newcommand{\Rres}[1]{| #1 \}}
\newcommand{\Lres}[1]{\{ #1 |}
\newcommand{\Def}{\stackrel{\rm def}{=}}
\newcommand{\Vc}{r^{-1}_{12}}
\newcommand{\vtwo}[2]{\left (\begin{array}{c} #1 \\ #2 \end{array} \right )}
\newcommand{\vtwob}[2]{\begin{array}{c} #1 \\ #2 \end{array}}
\newcommand{\cntr}[1]{{\rm con}(#1)}
\newcommand{\norm}[1]{\{#1\}}
\newcommand{\red}[1]{\{#1+\cntr{#1}\}}
\newcommand{\Half}{\frac{1}{2}}
\newcommand{\vect}[1]{{\bf #1}}
\newcommand{\threejm}[3]{\left ( #1~#2~#3\right )}
\newcommand{\twojm}[1]{\left ( #1 \right )}
\newcommand{\twoj}[1]{\{ #1 \}}
\newcommand{\sixj}[6]{\left \{ 
                       \begin{array}{ccc} #1 & #2 & #3\\
                                        #4 & #5 & #6
                      \end{array}
                     \right \}
                     }
\newcommand{\ninj}[9]{\left \{ 
                       \begin{array}{ccc} #1 & #2 & #3\\
                                          #4 & #5 & #6\\
                                          #7 & #8 & #9
                      \end{array}
                     \right \}
                     }
\newcommand{\threej}[3]{\{ #1 & #2 & #3 \}}
\newcommand{\abs}[1]{\left \arrowvert #1 \right \arrowvert }
\author{Chang-Kui Duan}
\affiliation{
Institute of Applied Physics and College of Electronic Engineering,
Chongqing University of Post and Telecommunications, Chongqing 400065, China.
}
\author{Michael F. Reid}
\affiliation{
Department of Physics and Astronomy and MacDiarmid Institute of
Advanced Materials and Nanotechnology, University of Canterbury, Christchurch,
  New Zealand}

\date{\today}

\begin{abstract}

The $4f-5d$ transition rates for rare-earth ions
in crystals can be calculated with an effective transition operator
acting between model $4f^N$ and $4f^{N-1}5d$ states calculated
with effective Hamiltonian, such as semi-empirical crystal
 Hamiltonian. The difference
of the effective transition operator from the original transition
operator is the corrections due to mixing in transition initial
and final states of excited configurations from both the center
ion and the ligand ions. These corrections are calculated
using many-body perturbation theory. For free ions, there
are important one-body and two-body corrections. The one-body
correction is proportional to the original
electric dipole operator with magnitude of approximately
$40\%$ of the uncorrected electric dipole moment.
Its effect is equivalent to scaling down the radial integral
$\ME {5d} r {4f}$, to about $60\%$ of the uncorrected HF value.
 The two-body correction has magnitude
of approximately $25\%$ relative to the uncorrected electric dipole moment.
For ions in crystals, there is an additional one-body
correction due to ligand polarization, whose magnitude is 
shown to be about $10\%$ of the uncorrected electric dipole moment.
\end{abstract}

\maketitle

\section{Introduction}

In the recent years, optical spectroscopy of lanthanides
ions in crystals involving $4f-5d$ transitions has been 
widely studied.  These transitions are electric dipole parity
allowed, and therefore suitable for efficient absorption of
VUV radiation from the noble gas discharge in mercury-free
lamps and plasma display panels.\cite{Bla1994,Ron1995,WegD1999}
 A well-known application
of the strong $4f-5d$ absorption of lanthanides is
in blue lamp phosphors BaMgAl$_{10}$O$_{17}$: Eu$^{2+}$. 
Potential applications for which VUV $5d-4f$ emission can be used are
tunable VUV lasers and scintillator materials. 

The transition rates are due to electric dipole
 moment contribution. The electric dipole matrix elements between
 $4f$ and $5d$ orbitals have been generally expected to dominate the
 electric dipole moment, and  most of the
 calculations\cite{ReiP2000,Dua2003,Col2003,Dzu2003} we are aware of
 consider only this zeroth-order contribution, which we shall refer to as the
uncorrected electric dipole moment in the paper. Recently, various detailed
 comparisons between theory and experiment for free ions
have shown that the zeroth-order calculations considerably
over-estimated the $5d\rightarrow 4f$ spontaneous emission rates
and configurational mixing needs to be included.\cite{Li2001,Bie2001,Zha2001}
Using the pseudorelativistic Hartree-Fock (HFR) approach of
Cowan\cite{Cow1981},
 a configuration interaction (CI) calculation for free-ion
 Ce$^{3+}$ includes both 12 odd-parity and  12 even-parity
 configurations gave radiative lifetimes agrees to experiment,\cite{Zha2001} 
 while a single configuration calculation gives uncorrected radial
 integral $\ME {5d}{r}{4f} = 0.0435 {\rm nm}$, which underestimates
 the lifetime, {\it i.\ e.}, overestimates the rates,  by a factor of about 3. For rare-earth ions 
in crystals, the calculation of electric-dipole allowed $4f-4f$
two-photon absorption rates has shown that some contributions due
to ligand excitations are very important.\cite{Bur2003} Since the ligand 
polarization contributions to two-photon transition moment are mainly due
to modification of one-photon transition moment between $4f^N$
 and $4f^{N-1}4d$ states, they are also potentially important to
$4f-5d$ one-photon transitions. 

Direct CI calculations similar to the one for
 Ce$^{3+}$ are less feasible for rare-earth ions with near 
half-filled $4f$ shell due to the large dimension of state spaces and 
become even more impractical when ligands orbitals are included, as
in the cases of quantum chemical cluster calculations.
A systematical method to take all those contributions into account
in a reduced space is effective operator method.\cite{Lin1984,Hur1993,DuaR2001}
In this method, the full Hilbert space time-independent
Hamiltonian $H$ is transformed into an effective Hamiltonian $H_{\rm
  eff}$, which acts on the reduced space (referred to as model space)
 and gives upon diagonalization a set of exact
eigenvalues and model space eigenvectors. For a time-independent
operator $O$, an effective
operator $O_{\rm eff}$ may be introduced that gives the same matrix
elements between the model space eigenvectors of $H_{\rm
  eff}$ as those of the original operator $O$ between the 
corresponding true eigenvectors of $H$. Effective Hamiltonians
 $H_{\rm eff}$ and transition operators $O_{\rm eff}$ 
are often constructed by many-body
perturbation theory (MBPT) with order by order
approximation\cite{DuaR2001} and then 
represented by connected diagrams similar to Feynman diagrams. With
those diagrams, perturbation calculations involving great number of
many-body intermediate states transform into calculations involving
summations over one- and two-particle states. Hence the calculations
are usually greatly simplified and the results are much easy to 
interpret since we can find out transition mechanism from diagrams
with important contributions. The calculations can also be easily used
together will semi-empirical crystal Hamiltonian for $4f^N$
configuration \cite{CarG1989} and its adaption to $4f^{N-1}5d$
configuration,\cite{ReiP2000} and with more general ligand field approaches.

In this paper we do a general many-body perturbation calculation 
for the $4f-5d$ effective transition operator.
The zeroth and first-order contributions
to the effective electric dipole operator are presented in Sec. II 
in terms of diagrams and algebra expressions. In Sec. III we
make further approximations to select out important contributions
and discuss the possible consequences to $4f-5d$
one-photon and $4f\rightarrow 4f$ two-photon transitions.

\section{General many-body perturbative expansion for effective
  operator and  diagrammatic representation}

We partition the Hamiltonian $H$ for the center-ion-ligand system
as usual into three terms, 
\be
H = H_0+V_1 + V_2
\ee
where $H_0$ is a zeroth order model Hamiltonian, $V_1$ 
and $V_2$ are the part of one-body spin-orbit and two-body
 Coulomb interactions, respectively,  not included 
in $H_0$. Usually one choose $H_0$ to be spherical
 and spin independent, here
we retain spin-independence but do not limit $H_0$
 to be spherical, so that if necessary
the strong crystal-field interactions for
 $5d$ electron in crystals can be included in $H_0$, as
in the calculation of $4f\rightarrow 4f$ two-photon absorption spectra in
Eu$^{2+}$:CaF$_2$.\cite{DuaR2002} In general, the effect of crystal-field
interactions to $f-d$ transitions is mainly through the effect to eigenstates
of the center ion. 

The eigenstates of $H_0$ are antisymmetric product
of one-particle states from the center ion and
ligand ions.  The model space is chosen to contain
the product of both $4f^N$ and $4f^{N-1}5d$ states with states of 
completely filled shells from the ligands.

The effective Hamiltonians can be calculated
with a many-body perturbation expansion.\cite{Bra1967}
However, for $4f^N$ energy levels of rare-earth ions,
the phenomenological crystal-field Hamiltonian\cite{New2000}
with adjustable parameters turns to produce better energy levels. 
The phenomenological crystal-field
Hamiltonian has also been adapted to the calculation of
$4f^{N-1}5d$ energy levels.\cite{ReiP2000}
 Since the eigenvectors of the phenomenological Hamiltonian can 
be used as the model space eigenvectors of the Hermitian
effective Hamiltonian, we need only construct
the Hermitian effective transition operator.

We consider transition due to electric dipole mechanism only.
The transition operator is:
\be 
{\bf D} = -e \sum\limits_{i=1}^{N} \vec{r_i}.
\ee
In addition to the zeroth-order contribution included in most
calculations, here we include also the first-order contribution.
Using our latest results on effective operators and modified
diagrams,\cite{DuaR2001,DuaR2004a}
we get the zeroth- and first-order diagrams for
the effective transition operator as shown in Fig.\ref{figure1}.

Since the effective operator acts 
between eigenvectors of effective Hamiltonian,
which are linear combinations of bases in model space,
 the incoming 
lines and outgoing lines in Fig.\ref{figure1} can only be $4f$ orbitals
and $5d$ orbitals.  The internal lines 
can be any orbitals of the center ion and ligands, as long
as at any horizontal level at least one of them is a core or
unoccupied orbital.

The evaluation rules have been summarized previously by
Duan {\it et al.}\cite{DuaR2004a}  For these diagrams
in Fig. \ref{figure1}, the contribution from
each diagram is proportional to the product of matrix
 elements of all vertexes divided by the
net outflow energy (outflow energy taking 
away inflow energy) of the $V$ vertex 
in the diagram.  The algebraic expressions are as follows:
\begin{eqnarray}
(a) &=& \sum\limits_{m,n} a_m^+a_n \cdot {\bf D}_{m,n}\\
(b) &=& \sum \limits_{m,n} a_m^+a_n \cdot 
              \sum\limits_i\frac{{\bf D}_{m,i}(V_1)_{i,n}}
                                   {\epsilon_n-\epsilon_i} \\
(c) &=& \sum \limits_{m,n}  a_m^+a_n \cdot 
                \sum\limits_i\frac{(V_1)_{m,i}{\bf D}_{i,n}}
                                   {\epsilon_m-\epsilon_i}\\
(d) &=& \sum \limits_{m,n}  a_m^+a_n \cdot 
                     \sum\limits_{u,a}
    \left [\frac{(V_2)_{mu,na}{\bf D}_{a,u}}
                            {\epsilon_n+\epsilon_a-\epsilon_m-\epsilon_u}
                      +\frac{(V_2)_{ma,nu}{\bf D}_{u,a}}
                            {\epsilon_m+\epsilon_a-\epsilon_n-\epsilon_u}
                      \right ]\nonumber\\
(e) &=& \frac{1}{4}\sum \limits_{mn,pq}  a_m^+a^+_na_qa_p \cdot 
                 \sum\limits_i \frac{2(V_2)_{mn,pi}{\bf D}_{i,q}}
                        {\epsilon_m+\epsilon_n-\epsilon_p-\epsilon_i}\\
(f) &=&\frac{1}{4} \sum \limits_{mn,pq} a_m^+a^+_na_qa_p \cdot 
                 \sum\limits_i \frac{2(V_2)_{mi,pq}{\bf D}_{n,i}}
                        {\epsilon_p+\epsilon_q -\epsilon_m-\epsilon_i},
\end{eqnarray}
where $m,~n,~p,~q$ are valence orbitals, {\it i.e.}, $f$ or $d$ orbitals,  $i$ 
can be any core or unoccupied orbital of the center ion or ligand ions,
 $u$ can be a valence or unoccupied orbital of the center ion
or an unoccupied orbital of the ligand ions,
$a$ is a core orbital of the center ion or ligand ions.
 Note that the two-body matrix elements are antisymmetric,
{\it i.e.}
\be
(V_2)_{ab,cd} = \ME{ab}{V_2}{cd} - \ME {ab}{V_2}{dc}.
\ee

\section{Further approximations and important contributions}

We make the following two approximations: (1)
we follow the usual approximation by neglecting 
integrals contain overlap of $4f$ orbitals of the
center ion with ligand orbitals; (2) we neglect
the extremely weak spin-orbit interactions between 
two orbitals $nl$ and $n^{\prime}l$ ($n\neq n^{\prime}$).\cite{Cow1981} With
these approximations, the one-body contributions and two-body
contributions can be greatly simplified and the strength can be 
calculated using structure data and HFR results for free ions.

\subsection{One-body contributions}

The zeroth-order term is the electric dipole between $4f$ and $5d$
orbitals, denoted by
Fig.\ref{figure1}(a). The first-order diagrams Fig.\ref{figure1}(b,c)
can be neglected with the second approximations above. This leaves
the only one-body contributions from Fig.\ref{figure1}(d). Under the 
first approximation above, there are two kind of contributions from
Fig.\ref{figure1}(d): (1) the contribution with the two internal lines
 are orbitals of the center ion, which does not change much from
free ion to ions in crystals; (2) the contribution with the two 
internal lines are orbitals of the ligands, which is usually denoted
 as ``dynamical coupling'' or ``ligand polarization'' contribution. 
The contribution due to ligand-center ion overlap are neglected
 under the first approximation since it contains $4f$-ligand overlap.

{\bf The first contribution.} For the case of free ions, 
since both the Coulomb interaction $V_2$
and the energy denominator in the expression for Fig.\ref{figure1}(d)
are spherically symmetric and spin-independent, the resulting
interaction for Fig.\ref{figure1}(d) can only be one-body
spin-independent rank-$1$ spherical tensor, which is proportional
to the electric dipole between $4f$ and $5d$ orbitals with opposite sign.
This provides a solid ground for calculations adopting an effective
radial matrix element $\ME {5d} r {4f}_{\rm eff}$ scaled
down from $\ME {5d}{r}{4f}$. 
This contribution is dominated by the terms
with the pair of internal lines ($5p$, $5d$), which have
small energy denominators and large matrix elements.
 After complex manipulation of
coupling and recoupling coefficients, we get the effective
operator in terms of radial integrals and (re)coupling coefficients
as follows:
\begin{widetext}
\begin{eqnarray}
&&{\bf D}_{\rm eff}[(d),1] = -\delta {\bf D} \\
&&~~~~~ = \delta (-e\ME {5d} r {4f}) (a_{5d}^+\tilde{a}_{4f})^1 +{\rm
  C.\ C.}\\
&&\delta = 
\sum\limits_{K,i_1,i_2}
 \frac{R^K(5di_1;4fi_2)}{
      \abs{\epsilon_{5d}+\epsilon_{i_1}-\epsilon_{4f}-\epsilon_{i_2}}}
 \frac{\ME {i_2} r {i_1}}{\ME {5d} r {4f}} 
\left [
\delta_{K1} \frac{\RME {i_2} {C^{(1)}} {i_1}^2}{3} 
\right .
\nonumber\\ && \left .
+ \sixj d 1 f {i_1} K {i_2} 
\frac{\RME {5d} {C^{(K)}} {i_2} \RME {i_2} {C^{(1)}} {i_1}
      \RME {i_1} {C^{(K)}} {4f}}{\RME {5d} {C^{(1)}} {4f}}
\right ],
\end{eqnarray}
\end{widetext}
where C.\ C. denotes the complex conjugative term,
$(i_1,i_2)$ is a pair of single-electron radial wavefunction
indices, with one being core orbitals and the other valence or
unoccupied orbitals. The leading terms are
\begin{widetext}
\begin{eqnarray}
\label{leadingterm}
&&\delta \approx
    \frac{R^{1}(5d5p;4f5d) \ME {5d} r {5p}} 
                 {\ME {5d} r{4f}}
  \left ( \frac{2}{3(\epsilon_{4f} - \epsilon_{5p})}
         +\frac{4}{15(2\epsilon_{5d}-\epsilon_{4f}-\epsilon_{5p})}
  \right )
\\ &&~~~
   +\frac{2R^{2}(5d5p;5d4f) \ME {5d} r {5p}}
                 {35 \ME {5d} r {4f}
                 (\epsilon_{4f}-\epsilon_{5p})}
   +\frac{2R^{4}(5d5p;5d4f) \ME {5d} r {5p}}
                 {21 \ME {5d} r {4f}
                 (\epsilon_{4f}-\epsilon_{5p})}
\\ &&~~~
 -\frac{4R^{3}(5d5d;5p4f) \ME {5d} r {5p}}
                 {35 \ME {5d} r {4f}
                 (2\epsilon_{5d}-\epsilon_{5p}-\epsilon_{4f})}.
\end{eqnarray}
\end{widetext}
The values of $\delta$ calculated from the leading terms are
$0.3571$ and $0.3502$ for Ce$^{3+}$ and Pr$^{3+}$ respectively.
Note that $\delta$ or {\bf D}$_{\rm eff}$ does not
depend on the crystal environment.

For free ion Ce$^{3+}$, since there is only one active electron
in model space $4f+5d$, the whole effective operator is strictly a 
one-body operator, which is approximately proportionally to the 
original electric dipole ${\bf D}$. 
The lifetime values from recent measurement and 
CI calculation \cite{Zha2001} can be used to work out
the effective $\ME {5d} r {4f}_{\rm eff} \approx 0.025
{\rm nm}$. Compared to the uncorrected HFR value 
$\ME {5d} r {4f} _{\rm eff } = 0.0435 {\rm nm}$,\cite{ReiP2002}  this
gives an experimental $\delta$ value for Ce$^{3+}$, which is
$(1 - \ME {5d} r {4f}_{\rm eff}/\ME {5d} r {4f})  \approx 0.43$.

{\bf The second contribution.} Neglecting the overlap between orbitals
of the center ion and those of the ligand ions, we can do a
bipolar expansion to the coulomb interaction. 
We obtain the same result as has been given earlier
 by Reid and Richardson\cite{ReiR1984} in calculation of 
two-photon transitions: 
\begin{widetext}
\begin{eqnarray}
{\bf D}_ {\rm eff}((d),2) &=&
 (-e) \sum\limits_{k=2,4,6}\ME {5d}
 {r^{k-1}}{4f}[(k)(2k-1)(2k+1)/3]^{1/2}\times
\nonumber\\
&&\sum\limits_L \bar{\alpha}_L (\epsilon_{5d}-\epsilon_{4f})
  R_L^{-(k+1)}[ C^k(L) C^{k-1(i)}]^{{\bf 1}}\\
&=& (-e\ME {5d} r {4f}) \sum\limits_{k=2,4,6} \sqrt{2k+1}
(A^k(a^+_{5d} \tilde{a}_{4f})^{k-1})^{\bf 1} + {\rm C.\ C.},
\\
A^k_q &=& \sqrt{k}\frac{\ME {5d} {r^(k-1)} {4f} \RME {5d} {C^{k-1}}{4f}}
             {\ME {5d} r {4f}}
         \sum\limits_{L}
          \frac{\bar{\alpha}_L (\epsilon_{5d} - \epsilon_{4f})C^k_q(L)}
               {R_L^{k+1}}.
\end{eqnarray}
\end{widetext}
where $L$ labels ligands, $i$ labels the valence rare-earth electrons, 
and 
\begin{widetext}
\begin{eqnarray}
\bar{\alpha}_L (\omega) 
 = \frac{1}{3}\sum\limits_{q} \sum\limits_{c,u} 
   \ME {|\phi_c} {r_q} {\phi_u}|^2 \left (
   \frac{1}{\epsilon_u-\epsilon_a+\omega}
   +\frac{1}{\epsilon_u-\epsilon_a-\omega} \right )
\end{eqnarray}
\end{widetext}
is the isotropic polarizability of ligand $L$. In the summation,
$q$ is over the component of $r_q$,
and $c$ and $u$ are over core and unoccupied
orbitals of ligand $L$, respectively.
Here we introduce dimensionless coefficients $A^k_q$ ( $k=2,4,6$, 
$q = -k,-k+1,\cdots, k$) with associated operators properly
normalized so that their values reflect the
relative magnitudes of the contributions compared to 
uncorrected electric dipole moment.

For Ce$^{3+}$: CaF$_2$, we use the structure data for CaF$_2$, setting 
$\bar{\alpha}_L \approx 10^{-3}  {\rm nm}^3$
for $F^{-}$,\cite{ReiR1984} and Ce$^{3+}$ free-ion data\cite{ReiP2002}
to obtain the $A^k_q$ values. The nonzero coefficients are $A^4_0 = \sqrt{14/5}
A^4_{\pm 4} = -0.08$ and $A^6_0 = -\sqrt{2/7} A^6_{\pm 4} = 0.0064$.
The case for $4f-5d$ transitions of Gd$^{3+}$:LaF$_3$ is estimated to be
similar. Therefore, corrections due to ligand polarization is not
so important for $4f-5d$ transitions as the corrections due to the
excited states of the
center ion. The reason that the ligand polarization contributions
are important for $^8S_{7/2}\rightarrow\ {^6I}_{J}$ $4f\rightarrow 4f$
 two-photon absorption is due to superposition of contributions from different
 intermediate states: the transitions are between states with main
components satisfying $\Delta L = 6 $, most part of the contribution
due to the $4f-5d$ rank-one uncorrected electric dipole contributions
 cancels, but the rank-3 and rank-5 ligand polarization contributions
 do not cancel and become important.

\subsection{Two-body contributions}

The two diagrams Fig.\ref{figure1}(e) and Fig.\ref{figure1}(f) are
complex conjugation of each other. To give a nonzero contribution, 
the single internal line can only be an orbital of the center ion due
 to the negligible overlap between $4f$ and ligand orbitals.
 These two diagrams are likely 
to be dominated by the two terms with $5p$ internal lines, which have
energy denominators and matrix elements comparable to the important
term in Fig.\ref{figure1}(d). The two terms with $5p$ internal lines,
which are complex conjugation of each other, can be written as
(with the complex conjugate term neglected)
\begin{widetext}
\begin{eqnarray}
&&{\bf D} _{\rm eff}[{(e,f)},0]
= \frac{1}{2} \sum\limits_{m_1,m_2,m_3,m_4} 
a_{4f m_1}^+a_{5d m_2}^+ a_{4f m_3} a_{4f
  m_4} \times \nonumber \\
&&~~~~~ \frac{\sum\limits_{m_5}\ME {4fm_15pm_5} {V_2} {4fm_3 4f_m4}
                                \ME {5d m_2} {{\bf D}} {5p m_5}
                              }{\epsilon_{4f} - \epsilon_{5p}} + {\rm C.C.}
\\
&&= (\epsilon_{4f}-\epsilon_{5p})^{-1} 
 \sum\limits_{m_1,m_2,m_3,m_4} a_{4f m_1}^+a_{5d m_2}^+ a_{4f m_3} a_{4f
  m_4} \times \nonumber\sum\limits_K R^K(4f5p;4f4f)  +{\rm C.C.}
\\
&&~~~~~ \sum\limits_{q m_5} (-1)^q \ME {4f m_1} {C^K_q}{4f m_3} 
                 \ME {5d m_2} {{\bf D}} {5p m_5} \ME {5p m_5} {C^K_{-q}}
                 {4f m_4} +{\rm C.C.} \\
&&= d(K,Q) (-e \ME {5d} r {4f}) \frac{[(a_{4f}^+ \tilde {a_{4f}})^K (a_{5d}^+
\tilde{a_{4f}})^Q ]^1}{2} +{\rm C.C.}
\label{two-body}
\end{eqnarray} 
\end{widetext}
where $K=2,4$, $Q =K-1$, $K$, and $K+1$ and the dimensionless 
coefficients
\begin{widetext}
\begin{eqnarray}
&&d (K,Q) = \frac{R^K(4f5p;4f4f)\ME
  {5d}{r}{5p}}{(\epsilon_{4f}-\epsilon_{5p})
  \ME {5d} r {4f}}
 \times \nonumber\\
&&\RME {4f}{C^K}{4f} \RME{5d}{C^1}{5p}\RME{5p}{C^K}{4f}(-1)^{K+Q+1} 
 2\sqrt{\frac{(2Q+1)}{3(2K+1)}} \sixj 1 1 2 3 Q K.
\end{eqnarray}
\end{widetext}
The operators associated with $d(K,Q)$ are appropriated normalized
so that the relative strengths can be reflected by these $d(K,Q)$
coefficients. The $d(K,Q)$ values for Pr$^{3+}$ are given in 
Table \ref{table1}. These values show that the corrections due
to two-body effective operators have a magnitude of $25\%$ and
can be important for rare-earth ions with two or more electrons
in $4f$ and $5d$ shells. Since in general matrix elements of 
two-body operators are not proportional to those of one-body 
operators, the effect due to two-body effective operators cannot
be fully accounted for by scaling the electric dipole operator.

All the different contributions to $4f-5d$ electric dipole transitions
we considered and their relative magnitude are listed in Table \ref{table2}. 
The uncorrected electric dipole term and one-body free-ion term 
are proportional to each other and form the main contribution
to $4f-5d$ transitions, which are equivalent to scaling the $4f-5d$
electric dipole radial integral with a factor $60\%$ ($= 1-40\%$). 
We call this corrected one-body free-ion term.
The next important contribution is the two-body free-ion term, which
has a relative magnitude of about $25\%$ of the uncorrected electric dipole
term and $40\%$ ($\approx 25\%/60\%$) of the corrected one-body free-ion
term. This term has the same rank as corrected one-body free-ion term (rank-1),
but is two-body and cannot be fully accounted for by scaling the
$4f-5d$ electric dipole radial integral. The ligand-polarization term
is about $10\%$ or the uncorrected electric dipole and $15\%$
of corrected one-body free-ion term. This term is less important
than two-body free-ion term for $4f-5d$ transitions but the high rank
(rank-3 and rank-5) part can be important for $4f-4f$ two-photon 
transitions with $\Delta L >2$.

\section{Summary}
In summary, we have presented a first-order many-body perturbation
calculation of the effective operator for $4f-5d$ transitions to account for
the main corrections to the transition rates due to mixing of other configurations
of the center ion and ligand orbitals in $4f^N$ and $4f^{N-1}5d$ states.
 Further approximations are made to select out possible 
important corrections to the usual zeroth-order $4f-5d$ 
electric dipole operator. First-order one-body contributions
due to excited states of the center ion is shown equivalent 
to scale down the electric dipole radial integral 
$\ME {5d} r {4f}$, which, for Ce$^{3+}$ and Pr$^{3+}$, has almost reduced the 
effective value to about $60\%$ of the uncorrected values.
 For rare-earth ions with two or more electrons in $4f$
 and $5d$ open shells, first-order two-body contributions
 also have important contributions, which cannot be accounted
for by scaling the electric dipole radial integral 
$\ME {5d} r {4f}$.  The magnitudes
have been calculated to be about $25\%$ relative to the
 uncorrected electric dipole moment for Pr$^{3+}$.
Contributions due to ligand polarization are rederived, which
are the same as given earlier by Reid and Richardson.\cite{ReiR1984}
The magnitude is about $10\%$ relative to the uncorrected electric
dipole moment for Ce$^{3+}$:CaF$_2$. The contribution due to ligand-center ion
overlap is negligible, since it contains the overlap between $4f$ and ligand
orbitals.

\section*{Acknowledgment}

C.\ K.\ D.\ acknowledges support of this work by the National
 Natural Science Foundation of China, Grant No. 10274079 and 10404040.
\newpage
\bibliography{fd5}

\newpage
\section*{Figures}
\begin{figure}[H]

\caption{Zeroth and First order diagrams for the effective
 electric dipole transition operator, where the free lines
 are $4f$ and $5d$ orbitals, the internal lines in 
 diagrams (b), (c), (e) and (d) are core or unoccupied orbitals,
 one of the two internal lines in (d) is core orbital and the other
 can be unoccupied or valence orbitals. The rules to evaluate these
 diagrams can be found in Ref.\onlinecite{DuaR2004a}.
\label{figure1}
}
\end{figure}
\newpage
\section*{Tables}
\begin{table}[H]
\caption{The values of dimensionless coefficients for Pr$^{3+}$ two-body
transition effective operators, where $d_{\rm tot} = (\sum
d(K,Q)^2)^{1/2}$ denote the total magnitude of two-body interactions.
\label{table1}
}

\end{table}

\begin{table}[H]
\caption{\label{table2}
List of different contributions to $4f-5d$ transitions and their relative magnitude
}
\end{table}

\section*{}
\newpage
\centerline{Figure \ref{figure1}, Duan and Reid, Journal of Chemical Physics}
\vskip 2cm
\includegraphics[width=11cm]
   {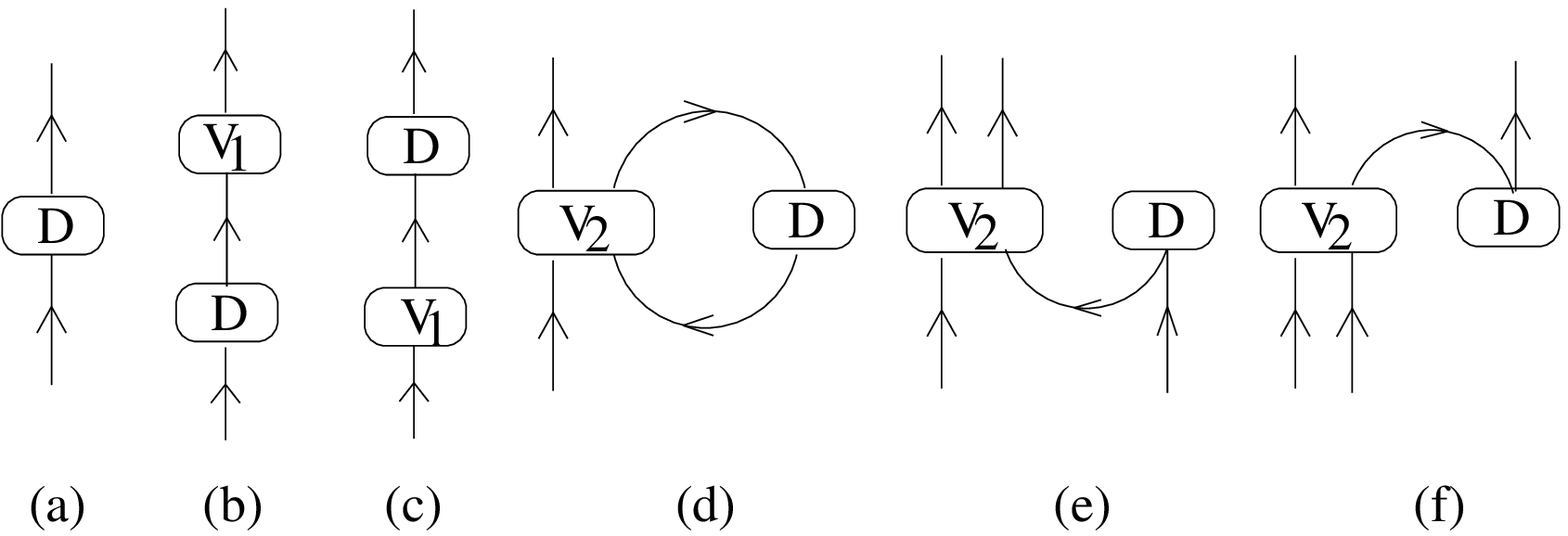}
\vskip 1cm

\newpage
\centerline{Table \ref{table1}, Duan and Reid, Journal of Chemical Physics}
\vskip 2cm

\begin{ruledtabular}
\begin{tabular}{ccccccc}
$~~d(21)~~$~~ & $d(22)$~~ & ~~$d(23)$~~ &$d(43)$ ~~& $d(44)$~~ & ~~$d(45)$~~ &~~$d_{\rm tot}$\\
-0.147  & -0.155  & -0.113  & -0.015 &  -0.035 &  -0.063 & 0.253
\end{tabular}
\end{ruledtabular}

\newpage

\centerline{Table \ref{table2}, Duan and Reid, Journal of Chemical Physics}
\vskip 2cm

\begin{ruledtabular}
\begin{tabular}{cccc}
uncorrected  &one-body free-ion & two-body free-ion &
ligand-polarization \\
  $1$          &   $(-) 40\%$     &    $25\%$         &
$10\%$
\end{tabular}
\end{ruledtabular}

\end{document}